\documentclass[11pt,a4paper]{article}
\usepackage{amsmath,amssymb,mathrsfs}
\usepackage[dvipdfmx]{graphicx}
\usepackage[usenames,dvipsnames]{color}

\newcommand{\vev}[1]{\left\langle #1\right\rangle}

\allowdisplaybreaks

\newcommand{\ms}[1]{\mathscr{#1}}
\newcommand{\mc}[1]{\mathcal{#1}}

\begin{document}

\begin{titlepage}

\begin{center}

{\small 
\hfill UT-14-46\\
\hfill RESCEU-49/14\\
\hfill WU-HEP-14-11}

\vskip .6in

{\huge \bf 
Reheating processes after Starobinsky inflation in old-minimal supergravity
}

\vskip .5in

{\large
Takahiro Terada$^{\spadesuit}$, 
Yuki Watanabe$^{\heartsuit}$, 
Yusuke Yamada$^{\diamondsuit}$,
Jun'ichi Yokoyama$^{\heartsuit , \clubsuit}$
}

\vskip 0.25in

{\em
${}^\spadesuit$Department of Physics, The University of Tokyo,
Tokyo 113-0033, Japan\\
${}^\heartsuit$Research Center for the Early Universe (RESCEU),
Graduate School of Science,\\
The University of Tokyo, Tokyo 113-0033, Japan\\
${}^\diamondsuit$Department of Physics, Waseda University, Tokyo 169-8555, Japan\\
${}^\clubsuit$Kavli Institute for the Physics and Mathematics of the Universe (Kavli IPMU),\\
WPI, TODIAS, The University of Tokyo, Chiba 277-8568, Japan
}

\end{center}
\vskip .5in

\begin{abstract}
We study reheating processes and its cosmological consequences in the Starobinsky model embedded in the old-minimal supergravity.
First, we consider minimal coupling between the gravity and matter sectors in the higher curvature theory, and transform it to the equivalent standard supergravity coupled to additional matter superfields.  We then discuss characteristic decay modes of the inflaton and the reheating temperature $T_{\rm R}$.  Considering a simple 
model of supersymmetry breaking sector, we estimate gravitino abundance from inflaton decay, and obtain limits on the masses of gravitino and supersymmetry breaking field.
We find $T_{\rm R}\simeq 1.0\times10^9$ GeV and the allowed range of  gravitino mass as $10^4$ GeV $\lesssim m_{3/2} \lesssim 10^5$ GeV, assuming anomaly-induced decay into the gauge sector as the dominant decay channel. 
\end{abstract}
\end{titlepage}

\section{Introduction}

The recent observations of cosmic microwave background (CMB) 
by the WMAP \cite{WMAP} and Planck satellites~\cite{Planck} 
indicate that nature is simple and minimal, 
providing increasing
evidence to favor  single field inflationary models.

Among the pioneering \cite{Starobinsky:1980te,sato,guth,Linde:1981mu, Albrecht:1982wi, Linde:1983gd} 
and newer models of inflation \cite{review}, 
Starobinsky model~\cite{Starobinsky:1980te} occupies a unique
position, since it does not require any new field
to drive inflation which we call the inflaton.  
Its original version
\cite{Starobinsky:1980te} is based on a higher curvature action which
emerges by incorporating matter loops to the Einstein Hilbert action.
Since all the second-order contributions of curvature tensors that
affect the Einstein equation in a conformally flat geometry including
the Robertson-Walker spacetime can be adequately described by the square
scalar curvature term, currently popular version of
Starobinsky model simply consists of linear and second-order terms of
the Ricci scalar $R$ \cite{Mijic:1986iv}.

In this model inflation is  followed by an
oscillatory behavior of the Hubble parameter, which results in
gravitational particle production to reheat the Universe.
Making use of a conformal transformation, one can also recast the system
to the Einstein action containing a scalar field, dubbed as the scalaron
acting as the inflaton, with a specific potential whose overall
magnitude is determined by the coefficient of $R^2$ term in the original
action~\cite{Whitt:1984pd, Maeda:1987xf}.  
In this picture reheating is described by the decay of the
scalaron. 
Various aspects of reheating after $R^2$ inflation have been
 studied in Refs.~\cite{Vilenkin:1985md, Watanabe:2006ku, Gorbunov:2010bn,
 Arbuzova:2011fu, Takeda:2014qma}. Ref.~\cite{Gorbunov:2010bn}
 showed that dark matter and baryon asymmetry are produced at 
reheating by introducing Majorana neutrinos.  Ref.~\cite{Takeda:2014qma} 
showed that parametric resonance is not strong enough to form
long-living 
localized objects, and thus reheating proceeds through perturbative 
decay of the inflaton.

The only  adjustable parameter of Starobinsky model, 
namely the coefficient of the curvature
square term, can be fixed by the
amplitude of curvature perturbation \cite{Mukhanov:1981xt}.
Its spectral index $n_s$ and the tensor-to-scalar ratio $r$ have also
been confronted with observations, and interestingly, this oldest
inflation model occupies just the central region of their error ellipses
\cite{WMAP,Planck}.  This feature was challenged this March by BICEP2
collaboration~\cite{Ade:2014xna}, which claimed to have detected 
B-mode polarization of CMB corresponding to a value
of $r$ much larger than favored by these satellite based observations.  
It turned out later, however,
 that the contamination of foreground dust may be so significant that
one cannot rule out models with small $r$ yet at all
~\cite{Mortonson:2014bja, Flauger:2014qra, Adam:2014bub}.
Thus the observational validity of Starobinsky model is still intact.

Needless to say, on the other hand, occupying the central region of the
likelihood contour does not necessarily mean the model is the right one,
and we should continue our efforts to further clarify  features of
Starobinsky model  in
the context of modern high energy theories, in particular, in
supersymmetry (SUSY) which reduces the hierarchy problem significantly,
naturally realizes gauge coupling unification, and provides
cold  dark matter candidates.

In the case of Starobinsky model, which is a theory of gravity, SUSY actually means supergravity (SUGRA)~\cite{Ref:SUGRA}.
There are two minimal choices for the SUGRA multiplet: the old-minimal~\cite{Ferrara:1978em, Stelle:1978ye, Fradkin:1978jq} and new-minimal~\cite{Sohnius:1981tp} formulations.
These two formulations utilize different SUGRA auxiliary fields, but they coincide on-shell for the standard SUGRA action, whose bosonic part is General Relativity.\footnote{In the case without any higher derivative terms, the equivalence between different formulations of SUGRA is shown from the conformal SUGRA viewpoint~\cite{Ferrara:1983dh}.}
This situation changes in the case of higher-derivative SUGRA, including SUGRA versions of Starobinsky model, because these auxiliary fields become propagating degrees of freedom.
Embedding of the Starobinsky action into the old-minimal SUGRA was studied at the linearized level in Ref.~\cite{Ferrara:1978rk} and the non-linear level in Ref.~\cite{Cecotti:1987sa}, where the duality to the standard SUGRA action with additional superfields was also established (in analogy with the bosonic case~\cite{Whitt:1984pd, Maeda:1987xf}).  Tachyonic instability during the inflationary phase was cured in Ref.~\cite{Kallosh:2013lkr}.
In this SUGRA setup, the $R+R^2$ action (without higher order terms) emerges from generic $F$- and $D$-term action (without derivatives and superderivatives)~\cite{Cecotti:1987sa, Hindawi:1995qa} (see also Refs.~\cite{Ketov:2013sfa, Ketov:2013dfa}).
The limited case, $F$-term generic action, was rediscovered in Ref.~\cite{Gates:2009hu} and developed, \textit{e.g.}, in Refs.~\cite{Ketov:2012yz, Ketov:2013wha} and 
its cosmological application was considered in Refs.~\cite{Ketov:2012se, Watanabe:2013lwa},
 but actually $D$-term action is required to realize Starobinsky inflation~\cite{eno, Ferrara:2013wka, Ketov:2013dfa, Ferrara:2013pla}.
In the new-minimal SUGRA, embedding of Starobinsky model was studied in Ref.~\cite{Cecotti:1987qe} and reconsidered in the inflationary context with higher order corrections in Ricci scalar in Ref.~\cite{Farakos:2013cqa}.  See also Ref.~\cite{fklp}.

Besides these pure SUGRA models without matter, there are many SUGRA models with matter that have Starobinsky-like scalar potentials (see, as an incomplete list, Refs.~\cite{Cecotti:1987sa, eno, Kallosh:2013lkr, SCDterm, Farakos:2013cqa, fklp, Ferrara:2013wka, Ketov:2013dfa, Alexandre:2013nqa, Pallis:2013yda}), so it is of prime importance to distinguish these models by studying cosmological scenarios after inflation.
To discuss reheating of the universe, one has to couple the pure SUGRA inflation sector to matter sector.
Ref.~\cite{Hindawi:1996qi} studied soft SUSY breaking pattern in the old-minimal case, whereas
Ref.~\cite{Ferrara:2014cca} discussed some features of matter-coupling in the new-minimal setup.

In this paper, we consider generic old-minimal SUGRA models~\cite{Cecotti:1987sa, Hindawi:1995qa, Ketov:2013dfa} that realize Starobinsky inflation focusing particularly on the model in Ref.~\cite{Kallosh:2013lkr}.
One of the reasons for the choice of the old-minimal formulation is that one has eventually to break R-symmetry to give gauginos their masses, but the new-minimal formulation has an exact R-symmetry.
We assume the absence of even higher order terms in scalar curvature, corresponding to absence of superderivatives in the SUGRA action, because such terms may modify or hamper inflation \cite{Kamada:2014gma}.
We introduce matter-coupling and study its cosmological consequences.
In particular, we study various inflaton decay channels extensively.
In contrast to the original non-SUSY version of reheating in the Starobinsky model, there is a long-lived particle, gravitino.
The gravitino is the superpartner of the graviton and hence always present in SUGRA, and its abundance is a cosmologically important subject.
We thus study the partial decay rate into gravitinos, and resultant constraints on parameters of the theory.

Before explaining our setup in section~\ref{sec:setup}, we briefly emphasize the differences from the literature.
In our setup, as we will see, the inflaton must have specific super- and K\"{a}hler potentials.
For example, the exponential of K\"{a}hler potential \textit{linearly} depends on the real part of the inflaton, while the gauge kinetic function never depends on the inflaton.
Our setup, the old-minimal SUGRA realization of the Starobinsky model, is thus predictive.
To the best of our knowledge, 
this is the first  study of inflaton decay and gravitino production in the theory described by a modified action of supergravity.
In section~\ref{sec:decay}, we study various partial decay rates of the inflaton.
In section~\ref{sec:cosmology}, we discuss the cosmological constraints from gravitino abundance.
We summarize and discuss differences from the original (non-SUSY) version of the Starobinsky model in section~\ref{sec:conclusion}.
The duality transformation between a higher derivative SUGRA and the corresponding standard SUGRA is reviewed, and some generalization of it is discussed in Appendix~\ref{app}.
We use the reduced Planck unit $c=\hbar=M_{\text{G}}=1$ with $M_{\text{G}}=M_{\text{Pl}}/\sqrt{8\pi} =1/\sqrt{8\pi G}$ unless otherwise stated,
and basically use the notation and convention of Ref.~\cite{Wess:1992cp}.

\section{Starobinsky model embedded in matter-coupled old-minimal supergravity} \label{sec:setup}
The Starobinsky model is based on a pure gravity action 
with a second order term of scalar curvature.
In the supergravity side, a generic (super)gravitational action up to matter and (super)derivatives is 
\begin{align}
S_{\rm grav}=& \int \mathrm{d}^4x\mathrm{d}^{4}\theta E  N(\mc{R},\bar{\mc R}) + \left[ \int \mathrm{d}^{4}x\mathrm{d}^{2}\Theta 2 \ms{E} 
 F(\mc{R})+\text{H.c.}\right] ,\label{minimal}
\end{align}
where $\mc{R}$ is the curvature chiral superfield, $E$ is the full density, $\ms{E}$ is the chiral density, $\Theta$ is the so-called new $\Theta$ variable~\cite{Wess:1992cp}, $N(\mc{R},\bar{\mc{R}})$ is a Hermitian function, and $F(\mc{R})$ is a holomorphic function.\footnote{The first, non-holomorphic term is called $D$-term action as it is from $D$-component of K\"ahler potential of ${\cal R}$, while the second, holomorphic term is called $F$-term action as it is from $F$-component of superpotential of ${\cal R}$.}

To discuss inflaton decay and reheating of the universe, we consider a simple way of coupling the above action to matter sector.  
We take the minimal coupling between the SUGRA sector described by the curvature chiral superfield $\mc{R}$ and the matter sector described by chiral superfields $\phi^{i}$ and vector superfields $V^{A}$:
\begin{align}
S=& \int \mathrm{d}^4x\mathrm{d}^{4}\theta E \left( N(\mc{R},\bar{\mc R}) +J\left(\phi,\bar{\phi}e^{gV}\right) \right) \nonumber\\
&+ \left[ \int \mathrm{d}^{4}x\mathrm{d}^{2}\Theta 2 \ms{E} 
\left( F(\mc{R})+P(\phi) + \frac{1}{4}h_{AB}(\phi)W^{A}W^{B} \right)+\text{H.c.}\right] \nonumber \\
=&\int \mathrm{d}^4x\mathrm{d}^{4}\theta E  N(\mc{R},\bar{\mc R}) \nonumber \\
& + \left[ \int \mathrm{d}^{4}x\mathrm{d}^{2}\Theta 2 \ms{E} 
\left( F(\mc{R})+\frac{3}{8}\left( \bar{\ms{D}}\bar{\ms{D}}-8\mc{R}\right)e^{-K^{(\phi)}/3}+P(\phi)+  \frac{1}{4}h_{AB}(\phi)W^{A}W^{B}\right)+\text{H.c.}\right] ,\label{OriginalTheory}
\end{align}
where $g$ is the gauge coupling constant, $\phi$ collectively denotes $\phi^i$'s, $J\left(\phi,\bar{\phi}e^{gV}\right)$ is a Hermitian function,  $P(\phi)$ is a holomorphic function, and $K^{(\phi)}(\phi,\bar{\phi}e^{gV})=-3\ln \left( -\frac{J(\phi,\bar{\phi}e^{gV})}{3} \right)$ is the K\"{a}hler potential of the matter fields.

The above action can be recast into the following form~\cite{Cecotti:1987sa, Ferrara:2014cca}:
\begin{align}
S=\int d^{4}xd^{2}\Theta 2\ms{E} \frac{3}{8}\left( \bar{\ms{D}}\bar{\ms{D}}-8\mc{R}\right)e^{-K/3}+W + \frac{1}{4}h_{AB}W^{A}W^{B}+\text{H.c.} ,\label{TransformedTheory}
\end{align}
with the K\"{a}hler potential and superpotential specified as follows,
\begin{align}
K &= -3\ln\left( \frac{T+\bar{T}-N(S,\bar{S})-J(\phi,\bar{\phi}e^{gV})}{3} \right) , \label{kahlerp} \\
W &=2TS+F(S)+P(\phi) \label{superp}.
\end{align}
The derivation (in a more general setup) is reviewed in Appendix~\ref{app}.
Note that the dependence of these potentials on the inflaton $T$ is completely determined by the structure of the theory: the origin of the inflaton $T$ is the Lagrange multiplier.\footnote{Recently, the work~\cite{Diamandis:2014vxa} suggested a higher derivative SUGRA model in which a superpotential term of $S$ and $T$ is given by $W=g(T)S$. Such a superpotential can be realized if $T$ is not a Lagrange multiplier but a chiral multiplet coupled to $\cal R$ and $\bar{\cal R}$ (see Ref.~\cite{Cecotti:2014ipa} for an earlier discussion). We briefly discuss similar extensions in Appendix~\ref{app}.
 In this work, we discuss the minimal case that the chiral multiplets $T$ and $S$ are purely originated from the gravitational multiplet and its higher derivative modes, and that the superpotential term of $T$ and $S$ is given by $W=2TS$ as in eq.~(\ref{superp}).}
This structure is not altered even if non-minimal couplings between ${\cal R}$ and matter superfields, which we do not discuss in this paper, are introduced because they become non-minimal couplings between $S$ (but not $T$) and matter superfields in the transformed theory.\footnote{We briefly discuss a possibility of $T$ dependent gauge kinetic functions in Appendix~\ref{app}.} 
Therefore, in this sense, the couplings between $T$ and matters discussed in this paper are universal in old-minimal Starobinsky inflation. 

The K\"{a}hler metric and its inverse are given by
\begin{align}
g_{I\bar{J}}=&\frac{3}{\left( T+\bar{T}-N-J \right)^{2}}\begin{pmatrix} 1 & -N_{\bar{S}} & -J_{\bar{j}} \\
-N_{S} & N_{S\bar{S}}\left( T+\bar{T}-N-J \right)+N_{S}N_{\bar{S}} & N_{S} J_{\bar{j}} \\
-J_{i} & N_{\bar{S}}J_{i} & J_{i\bar{j}}\left( T+\bar{T}-N-J \right)+J_{i}J_{\bar{j}} \end{pmatrix}, \\
g^{\bar{I}J}=&\frac{T+\bar{T}-N-J}{3} \begin{pmatrix} \left(T+\bar{T}-N-J\right) +N_{S}N^{S}+J_{k}J^{k} & N^{S} & J^{j} \\
N^{\bar{S}} & N^{\bar{S}S} & 0 \\
J^{\bar{i}} & 0 & J^{\bar{i}j} \end{pmatrix},
\end{align}
where $I,J,\dots=T, S, i,j,\dots \text{(or }\phi^{i},\phi^{j},\dots\text{)}$ are field indices, $N^{\bar{S}S}=(N_{S\bar{S}})^{-1}$, $J^{\bar{i}j}$ is the inverse matrix of $J_{i\bar{j}}$, and indices are uppered and lowered by these matrices, {\it e.g.} $N^{S}=N^{\bar{S}S}N_{\bar{S}}$ and $J^{\bar{i}}=J^{\bar{i}j}J_{j}$.
The scalar potential is
\begin{align}
V=&\left(\frac{3}{A}\right)^{2}\left( N^{\bar{S}S}\left| 2T+F_{S} \right|^{2}+\left| 2S \right|^{2}\left( A+N_{S}N^{S}+J_{i}J^{i} \right)+\bar{P}_{\bar{i}}J^{\bar{i}j}P_{j}  \right. \nonumber \\
& \left. \phantom{N^{\bar{i}j} \bar{P}_{\bar{i}}P_{j} } +\left \{ 2\bar{S} \left [ \left(2T+F_{S} \right) N^{S}+P_{i}J^{i}-3W \right ] + \text{h.c.} \right \}  \right) + \frac{g^{2}}{2}D^{A}D_{A} ,
\end{align}
where we have defined a compact notation $A\equiv T+\bar{T}-N-J$.\footnote{
It is often denoted as $\Omega=-3A$ in the standard notation~\cite{Wess:1992cp}, and $\tilde{\phi}=-3A$ in the conformal SUGRA notation~\cite{Kugo:1982mr}. The functional form of $\Omega$ is important for the SUSY breaking effects on inflationary dynamics~\cite{Abe:2014opa}. 
}  Indices of D-terms, $D^{A}\, (D_{A})$, are lowered (lifted) by (the inverse of) the real part of the gauge kinetic matrix function $h^{R}_{AB} \, (h_{R}^{AB})$.

The inflaton (or SUGRA) sector ($T$ and $S$) of this class of modified SUGRA models was studied in Ref.~\cite{Ketov:2013dfa}.
The Starobinsky model is realized in this setup essentially as the modified Cecotti model~\cite{Kallosh:2013lkr}:
\begin{align}
N(S,\bar{S})=&-3+\frac{12}{m_{\Phi}^{2}}S\bar{S}-\frac{\zeta}{m_{\Phi}^4} \left( S\bar{S} \right)^{2} \label{simpleN}, \\
F(S)=&0 \label{simpleF},
\end{align}
where $m_{\Phi}$ is the inflaton mass at the vacuum, and $\zeta$ $(>0)$ gives 
a SUSY-breaking mass to $S$ and stabilizes its potential.
The real part of $T$ becomes the inflaton, and the canonically normalized scalar potential is that of the Starobinsky model, $V=\frac{3m_{\Phi}^2}{4}\left(1-e^{-\sqrt{2/3}\widehat{\text{Re}T}}\right)^{2}$, where $\widehat{\text{Re}T}\equiv -K/\sqrt{6}$ is the canonically normalized inflaton field (during inflation).
$S$ is the sGoldstino field that breaks SUSY during inflation.
At the vacuum ($T=S=0$), SUSY is preserved.

Introduction of the linear term in $S/m_{\Phi}$ into eq.~\eqref{simpleN} can make SUSY breaking vacua with an almost vanishing cosmological constant without spoiling inflation~\cite{Dalianis:2014aya}.
This is an interesting possibility because the higher derivative version of the purely supergravitational theory describes not only the inflation but also SUSY breaking.
However, the SUSY breaking scale becomes the inflation scale ($m_{\Phi}\sim 10^{13}$ GeV), which typically makes the Higgs particle too heavy~\cite{Giudice:2011cg}.  Although the tree-level contributions to soft SUSY breaking parameters can be suppressed by assuming a minimal coupling between the MSSM sector and the SUGRA sector as in our setup, there are anomaly-mediated contributions to gaugino masses, which in turn give other particles their masses through renormalization group running.

Therefore, we concentrate on models that deviate (if any) only slightly from the simple model \eqref{simpleN}, \eqref{simpleF}.
For definiteness, we assume $|N^{S}|$ and $|N^{\bar{S}S}F_{SS}|$ are at most of order the gravitino mass $m_{3/2}$, which is supposed to be much smaller than the inflaton mass, $m_{3/2}\ll m_{\Phi}$. 
Perturbation by higher order terms are negligible because VEV of $S$ is suppressed.\footnote{
Although it vanishes at the leading order, it has a value of the order of the gravitino mass after SUSY breaking.  See the following discussion.
} 
Since the inflaton sector does not break SUSY at the vacuum, we introduce a hidden SUSY breaking sector.
We treat the SUSY breaking sector as general as possible, but occasionally we assume a simple SUSY breaking sector described by
\begin{align}
J(z, \bar{z})=& |z|^2 - \frac{|z|^4}{\Lambda^2} ,  \label{ZKahler}  \\
P(z)=& \mu^2 z +W_0 , \label{Zsuper}
\end{align}
where $J(z,\bar{z})$ and $P(z)$ are the K\"{a}hler potential and superpotential of the SUSY breaking field $z$ [see equations~\eqref{kahlerp} and \eqref{superp}].
We also assume that VEVs of $\phi^{i}$, $J(\phi^{i},\bar{\phi}^{\bar{j}})$, $P(\phi^{i})$, and their derivatives are negligibly small except for those of SUSY breaking field $z$, which is easily satisfied if $\phi^{i}$'s are charged under some unbroken symmetry.

All of the four scalar degrees of freedom and four fermionic degrees of freedom in the inflaton sector are degenerate in their masses ($=m_{\Phi}$) at the zeroth order of perturbation with respect to SUSY breaking ($m_{3/2}$).
In the scalar sector, imaginary parts of $T$ and $S$ are still degenerate at the first order of gravitino mass, but the sum and difference of real parts of $T$ and $S$ have mass eigenvalues $m_{\Phi}\mp m_{3/2}$.
Also, $S$ acquires its VEV, $\vev{S}=\vev{W}/2$.
Here we have neglected supersymmetric mass term of $S$ from its superpotential, $F_{SS}$.
Fermionic mass eigenvalues depend on the detail of functions $N$ and $F$, but in the simplest case \eqref{simpleN}, \eqref{simpleF}, they are still degenerate at the first order in gravitino mass.
For this kinematical reason, the decay of inflaton into particles in the inflaton sector (inflatino and gravitino), if possible, is extremely suppressed.

The mass eigenstates of the canonically normalized scalar linear
fluctuations are approximately given by 
\begin{align}
\Phi_{R\pm}=&\frac{\sqrt{g_{T\bar{T}}}}{2} \left(T+\bar{T}\right) \pm \frac{\sqrt{g_{S\bar{S}}}}{2} \left( S+\bar{S} \right) \simeq \frac{1}{2\sqrt{3}}\left( T+ \bar{T}\right) \pm \frac{\sqrt{3}}{m_{\Phi}}\left( S+\bar{S}\right). \label{MassEigenstate}
\end{align}
Because the $T$-$S$ oscillation time scale $\tau_{\text{osc}}\sim (2m_{3/2})^{-1}$ is much shorter than the lifetime $\tau_{\text{dec}}\sim (M_{\text{G}}^{2}/ m_{\Phi}^{3})$ for gravitino mass above GeV scale, decay rates from these mass eigenstates are appropriate quantities.
However, the interactions are simply described in the basis of $T$ and $S$ but not of their linear combination, so for simplicity of presentation we describe partial decay rates of inflaton in the next section as if $T$ (or $S$) is the parent particle.  The true rates are the averages of those for $T$ and $S$.

We take gravitino mass larger than TeV scale because we assume anomaly (or gravity) mediation of SUSY breaking,  in which SUSY breaking is transmitted to the visible sector  by the Planck suppressed coupling to the auxiliary field of the curvature superfield $\mc{R}$ in the transformed theory~\eqref{TransformedTheory} due to the trace anomaly (or by the Planck suppressed coupling to the hidden sector in the tree-level potential).
We respect the philosophy of the Starobinsky model in this paper, that is, we exploit the (super-)gravitational sector as much as possible, and do not introduce an inflaton nor messenger fields by hand.

\section{Inflaton decay} \label{sec:decay}
Various modulus/inflaton decay modes and their cosmological consequences have been extensively studied in Ref.~\cite{Endo:2007sz}.
Inflaton decay in the case of no-scale supergravity 
has also been  studied in Ref.~\cite{Endo:2006xg}, but in our case inflaton has supergravitational origin so that the form of inflaton K\"{a}hler potential is different from that in Ref.~\cite{Endo:2006xg}.
Moreover, these works suppose that the inflaton mass $m_{\Phi}$ comes mainly from the second derivative $W_{\Phi\Phi}$ of the superpotential with respect to the inflaton $\Phi$ itself.
In our case, on the other hand, the origin of the inflaton mass $m_{T}(\equiv m_{\Phi})$ is from $W_{TS}$ rather than $W_{TT}$. 
We study inflaton decay in our setup taking these differences into account.

At the end of inflation, the inflaton oscillates around the minimum of the potential for a long time due to its Planck-suppressed decay rate.
We have numerically checked that the energy stored in $\text{Re} T$ does not flow into $\text{Im} T$ or $S$ fields in this classical oscillation dynamics.
In the following, we study various partial decay rates of the inflaton at the tree-level unless the one-loop process becomes leading.
As stated at the end of the previous section, we first consider interactions involving $T$, followed by similar analyses for $S$.

\subsection{Two-body decay of $T$ into scalars, spinors and gauge bosons}

\subsubsection{Decay into scalars}

It is convenient to define the reduced scalar potential $\tilde{V}$ as $V=\left(\frac{3}{A}\right)^{2}\tilde{V}+\frac{g^{2}}{2}D^{A}D_{A}$, or equivalently,
\begin{align}
\tilde{V}=& N^{\bar{S}S}\left| 2T+F_{S} \right|^{2}+\left| 2S \right|^{2}\left( A+N_{S}N^{S}+J_{i}J^{i} \right)+\bar{P}_{\bar{i}}J^{\bar{i}j}P_{j} \nonumber\\
& +\left \{ 2\bar{S} \left [ \left(2T+F_{S} \right) N^{S}+P_{i}J^{i}-3W \right ] + \text{h.c.} \right \}.
\end{align}
Although $T$ and $S$ are singlets, derivatives of the $D$-term with respect to them are nonzero, 
\begin{align}
D_{AT}=&-ig_{T\bar{i}}\bar{X}_{A}^{\bar{i}}= \frac{3}{A^{2}}iJ_{\bar{i}}\bar{X}_{A}^{\bar{i}}=-\frac{1}{A}G_{\bar{i}}D_{A}{}^{\bar{i}}=-\frac{1}{A}D_{A}\simeq -\frac{1}{3}D_{A}, \\
D_{AS}=&-ig_{S\bar{i}}\bar{X}_{A}^{\bar{i}}= -\frac{3}{A^{2}}iN_{S}J_{\bar{i}}\bar{X}_{A}^{\bar{i}}=\frac{1}{A}N_{S}G_{\bar{i}}D_{A}{}^{\bar{i}}=\frac{1}{A}N_{A}D_{A}\simeq \frac{1}{3}N_{S}D_{A},
\end{align}
where $G=K+\ln{|W|^2}$ is the total K\"ahler potential, $X_{A}$ is the Killing vector of the K\"{a}hler manifold, and we have used the gauge symmetry of the superpotential.
With the aid of the condition of the vanishing cosmological constant, $V=0$, the stationary conditions for $T$ and $S$ at the vacuum, $V_{T}=V_{S}=0$, reduce to $\tilde{V}_{T}=\tilde{V}_{S}=0$.

Using the above formulas and the facts $\tilde{V}_{TT}=\tilde{V}_{Ti}=\tilde{V}_{T\bar{i}}=0$,
the relevant vertex functions are derived as
\begin{align}
V_{\tilde{T}\tilde{i}\tilde{j}}=-\frac{2}{A}V_{\tilde{i}\tilde{j}} \simeq -\frac{2}{3}V_{\tilde{i}\tilde{j}},
\end{align}
where tilded indexes may take both of holomorphic and anti-holomorphic indexes like $\tilde{I}=I, \bar{I}$.
This means that the interaction terms are proportional to the mass terms of scalars. 
There is a same order contribution from the kinetic term.
Combining mass and kinetic term contributions,  the rate is 
\begin{align}
\Gamma ( T \rightarrow \phi^{i}\bar{\phi}^{\bar{i}} ) = \frac{3m_{i}^{4}}{8\pi M_{\text{G}}^{2}m_{\Phi}},
\end{align}
where $m_{i}$ is the mass of the daughter particle $\phi^{i}$.
The kinetic term also provides the $\phi^{i}\phi^{j}$ production process with the rate
\begin{align}
\Gamma (T \rightarrow \phi^{i}\phi^{j}) = \frac{m_{\Phi}^{3}}{96\pi M_{\text{G}}^{2}}|J_{ij}|^{2}.
\end{align}

The partial decay rates of inflaton into $\text{Im}T$, $S$, or $\bar{S}$ and $\phi^{i}$ are suppressed by $J_{i}$ and phase space factors.

\subsubsection{Decay into spinors}

It is convenient to define the reduced fermion mass matrix $\tilde{M}$ as $M_{IJ}=e^{G/2}\tilde{M}_{IJ}$, where $M_{IJ}$ is the fermion mass matrix, or equivalently,
\begin{align}
\tilde{M}_{IJ}=\nabla_{I}G_{J}+G_{I}G_{J}-\frac{2}{3}\left( \langle G_{I} \rangle G_{J} + \langle G_{J} \rangle G_{I}  \right)+\frac{2}{3}\langle G_{I}G_{J} \rangle.
\end{align}
Terms with VEVs are induced by the redefinition of the gravitino field to absorb goldstino into gravitino.
The inflaton-spinor-spinor vertex is obtained by differentiating the mass matrix, $M_{IJT}=G_{T}M_{IJ}/2 + e^{G/2}\tilde{M}_{IJT}\simeq m_{3/2}\tilde{M}_{IJT}$.
Under the approximation like $A\simeq 3$ and $S\simeq W/2$, and neglecting $G_{i}, \, G_{T}$ and $G_{S}$,
the reduced fermion matrix $\tilde{M}_{ij}$ is approximated as $\tilde{M}_{ij}\simeq P_{ij}/W +J_{ij}-J_{ij\bar{z}}G^{\bar{z}}$ where $z$ is the SUSY breaking field.
Under the same approximation, 
\begin{align}
\tilde{M}_{ijT}\simeq -\tilde{M}_{ij}.
\end{align}
On the other hand, $\tilde{M}_{ij\bar{T}}$ vanishes at the vacuum.
The kinetic term gives a same order contribution.
Combining the mass and kinetic term contributions, the partial decay rate is expressed as
\begin{align}
\Gamma (T \rightarrow \chi^{i}\bar{\chi}^{\bar{i}} ) = \frac{m_{i}^{2}m_{\Phi}}{192\pi M_{\text{G}}^{2}},
\end{align}
where $m_{i}$ is the mass of the spinor $\chi^{i}$.
We have assumed here that the mixing terms between matter spinors and gauginos are smaller than the diagonal parts, $|M_{IA}|\ll |M_{JK}|$.

The partial decay rates of inflaton into inflatino or $S$-ino and $\chi^{i}$ are suppressed by $J_{i}$ and phase space factor.

\subsubsection{Anomaly-induced decay into gauge sector} \label{sec:anomaly}
The inflaton $T$ has the Lagrange multiplier origin
so that it never appears in the gauge kinetic function.  
We have to consider decay into gauge sector via the anomaly-induced one loop process~\cite{Endo:2007ih, Endo:2007sz} unless
we introduce a non-minimal term depending on $W^A$ in the $D$-term
action (see Appendix~\ref{app}).
The rate is~\cite{Endo:2007ih, Endo:2007sz}
\begin{align}
\Gamma (T \rightarrow AA) + \Gamma (T \rightarrow \lambda\lambda )\simeq \frac{N_{\text{g}}\alpha^{2}}{256\pi^{3}}|X_{G}|^{2}m_{\Phi}^{3},
\end{align}
where $N_{\text{g}}$ and $\alpha$ are the number of the generators and the fine structure constant of the gauge group, $X_{G}=\sqrt{6}\left [ (T_{G}-T_{R})K_{T}+\frac{2T_{R}}{d_{R}}\left( \log \det K|_{R}'' \right), _{T} \right ] $, $T_{G}$ and $T_{R}$ are the Dynkin indexes of the adjoint representation and representation $R$, $d_{R}$ is the dimension of the representation $R$, and $K|_{R}''$ is the K\"{a}hler metric restricted to the matter whose representation is $R$.
In our case, the rate becomes (also see \cite{Watanabe:2010vy} for non-SUSY case)
\begin{align}
\Gamma (T \rightarrow AA) + \Gamma (T \rightarrow \lambda\lambda )\simeq \frac{3N_{\text{g}}\alpha^{2}m_{\Phi}^{3}}{128\pi^{3}M_{\text{G}}^{2}}\left( T_{G}-\frac{1}{3}T_{R}\right)^{2}.
\end{align}

\subsection{Three-body decay of $T$}

Let us first consider the decay channel into a scalar and two spinors involving Yukawa coupling.
There are three diagrams at the tree level that are of the same order.
The effective interaction term that reproduces the decay rate is found to be~\cite{Endo:2007sz}
\begin{align}
\mc{L}_{\text{eff}}\simeq -\frac{1}{2}e^{G/2}\left( G_{Tijk}-3\Gamma _{T(i}^{l}G_{jk)l} \right)T\phi^{i}\chi^{j}\chi^{k}+\text{h.c.}
\end{align}
In our case,
the leading terms, which could lead to the typical Planck-suppressed decay rate, cancel each other, and the remaining terms give at most $\Gamma\sim m_{3/2}^{2}m_{\Phi}^{3}/M_{\text{G}}^{4}$.

There are also scalar three-body decay.
At the vacuum, the scalar four-point vertex is given by
\begin{align}
V_{\tilde{T}\tilde{i}\tilde{j}\tilde{k}}\simeq & -\frac{2}{3}\left( \tilde{V}_{\tilde{i}\tilde{j}\tilde{k}}+V_{D\tilde{i}\tilde{j}\tilde{k}}\right) .
\end{align}
The leading terms in $\tilde{V}_{ijk}$ cancel each other in the same way as for the above fermion case.
\begin{align}
\tilde{V}_{ijk}\simeq &  -3\bar{P}^{z}J_{z\bar{m}(i}J^{\bar{m}l}P_{jk)l} ,\\
\tilde{V}_{ij\bar{k}}\simeq & \bar{P}_{\bar{l}\bar{k}}J^{\bar{l}l}P_{ijl}-\bar{P}^{m}J_{m\bar{n}\bar{k}}J^{\bar{n}l}P_{ijl}+2\bar{S}P_{ijl}J^{\bar{l}l}\left( J_{\bar{l}\bar{k}}-J_{\bar{l}m\bar{k}}J^{m}\right) .
\end{align}
The rates are suppressed by gravitino or matter mass squared, $\Gamma \sim m_{X}^{2}m_{\Phi}/M_{\text{G}}^{2}$ with $m_{X}=\max [ m_{\text{(matter)}}, \, m_{3/2} ]$ at most.

We also considered decay modes involving $\text{Im}T$, $S$, or their superpartners, but these rates are at most of order of $m_{\Phi}^{5}/M_{\text{G}}^{4}$ with additional phase space suppression.  Four- or more- body decay rates are more suppressed by the phase space factor.

\subsection{Decay of $S$}

In the same way as the previous subsections, we study decay channel of $S$ in this subsection.
Although $S$ is basically conformally sequestered from the matter sector in our setup,
it has unsupressed coupling with $T$ in the superpotential, which in turn couples to the matter sector universally.  Consequently, $S$ has unsuppressed coupling to matter in some decay channels.
Important partial decay rates are as follows,
\begin{align}
\Gamma ( S \rightarrow \phi^{i}\phi^{i} ) \simeq & \frac{m_{i}^{2}m_{\Phi}}{48\pi M_{\text{G}}^{2}}, \\
\Gamma ( \bar{S} \rightarrow \chi^{i}\chi^{j} ) \simeq & \frac{m_{\Phi}^{3}}{48\pi M_{\text{G}}^{2}} \left| J_{ij} \right|^2 
\end{align}
Beware $m_{S}= m_{\Phi}$.
The above rates are calculated expanding mass terms.
If there are no heavy matter particles, the following contribution from kinetic term becomes important, 
\begin{align}
\Gamma (S \rightarrow \phi^{i}\phi^{j} )\simeq \frac{m_{\Phi}m_{3/2}^{2}}{192\pi M_{\text{G}}^{2}}|J_{ij}|^{2},
\end{align}
while other channels $\Gamma (S \rightarrow \phi^{i}\bar{\phi}^{\bar{j}})$ and $\Gamma (S \rightarrow \chi^{i}\chi^{j})$ from kinetic terms are suppressed by both of $N_{S}$ and matter masses.
For decay modes of $S$ involving $T$, see the previous subsections,  $\Gamma (S\rightarrow T X) = \Gamma (T\rightarrow S X)$.
The anomaly-induced decay of $S$ involves an additional $-N_{S}$ factor compared to the case of $T$.

\subsection{Gravitino production}
In this subsection we study gravitino production from the inflaton decay,  which is one of the distinguishing features from the non-SUSY version of the Starobinsky model.
Although we have treated $T$ or $S$ as the parent particle in the previous subsections, the mixing effect is essential in gravitino production~\cite{Dine:2006ii, Endo:2006tf}.
We thus use the proper mass eigenstates~\eqref{MassEigenstate} in evaluating the inflaton decay rate into gravitinos.

\subsubsection{Single gravitino production}

The partial decay rate of a scalar particle into its superpartner and a gravitino is calculated, {\it e.g.}, in Ref.~\cite{Buchmuller:2004rq}.  Because inflaton and inflatino are degenerate before SUSY breaking, their mass splitting is of the order of gravitino mass.
We parametrize the mass difference as
$	m_{\Phi}-m_{\tilde{\Phi}} = \Delta m_{3/2} $.
	The decay rate is approximately 
	\begin{align}
	\Gamma (\Phi \rightarrow \tilde{\Phi}\psi_{3/2} ) \simeq \frac{m_{\Phi}^{3}}{3\pi M_{\text{G}}^{2}}\left( \frac{m_{3/2}}{m_{\Phi}} \right)^{2}\Delta (\Delta^{2}-1)^{\frac{3}{2}},
	\end{align}
	where we explicitly wrote the reduced Planck mass $M_{\text{G}}$. Thus, the single gravitino production has the suppression factor $(m_{3/2}/m_{\Phi})^{2}$ and $(\Delta^{2}-1)^{3/2}$ compared to the typical Planck-suppressed decay rate $\mathcal{O}(m_{\Phi}^{3}/M_{\text{G}}^{2})$.
In subsequent discussion, we neglect the single gravitino production rate because it is at most of order of gravitino pair production rate discussed below.

\subsubsection{Gravitino pair production}
Gravitino pair production rate from modulus/inflaton decay has been extensively studied in the literature~\cite{Endo:2006zj, Nakamura:2006uc, Dine:2006ii, Endo:2006tf, Endo:2012yg}.  See also Refs.~\cite{Kawasaki:2006gs, Kawasaki:2006hm, Endo:2006qk, Asaka:2006bv, Endo:2007ih, Endo:2007sz} for other decay channels and cosmological consequences.

The gravitino pair production rate from a mass eigenstate $\Phi$ is given by \cite{Endo:2006zj, Nakamura:2006uc, Dine:2006ii, Endo:2006tf, Endo:2012yg}
\begin{align}
\Gamma(\Phi \to \psi_{3/2}\psi_{3/2}) =
\frac{|\mathcal{G}^{\text{(eff)}}_{\Phi}|^{2}m_{\Phi}^{5}}{288 \pi m^{2}_{3/2 }}, 
\label{gravitinorate}
\end{align}
where the mass hierarchy $m_{\Phi}\gg m_{3/2}$ is assumed, and the effective coupling is given by~\cite{Endo:2012yg}
$
\left | \mathcal{G}^{\text{(eff)}}_{\Phi} \right |^{2}
=2\left| G_{I} (\mc{A}^{-1}) ^{I}{}_{ \Phi} \right|^{2}, 
$
where $\mc{A}$ is the mixing matrix~\cite{Endo:2012yg}.
In our case, the inflaton is the real part of $T$,  
 but the real parts of $T$ and $S$ mix almost maximally at the vacuum (see eq.~\eqref{MassEigenstate}).

Because the SUSY breaking of $T$ and $S$ are small, $|G_{T}|, |G_{S}| \ll 1$, the effective coupling reduces to \begin{align}
\left | \mathcal{G}^{\text{(eff)}}_{\Phi_{\text{R}\pm}} \right |^{2}=2\left| \frac{\sqrt{3}}{2} G_{T} \pm \frac{m_{\Phi}}{4\sqrt{3}}G_{S}  +(\mc{A}^{-1})^{i}{}_{\Phi_{\text{R}\pm}}G_{i} \right|^{2}.
\end{align}
 We will first evaluate $G_T$ and $G_{S}$, and then proceed to $(\mc{A}^{-1})^{i}{}_{\Phi_{\text{R}\pm}}$.
We evaluate $G_{T}$ using the conditions $V=e^{G}(G_{I}G^{I}-3)+(g^{2}/2)D^{A}D_{A}=0$ for the vanishing cosmological constant and $V_{\bar{I}}=e^{G}(G_{\bar{I}}G^{J}G_{J}-2G_{\bar{I}}+G^{\bar{J}}\nabla_{\bar{I}}G_{\bar{J}})+g^{2}(-(h^{R}_{ABI}/2)D^{A}D^{B}+D^{A}D_{AI})=0$, where $\nabla_{I}G_{J}=G_{IJ}-G_{IJ\bar{K}}G^{\bar{K}}$, for the stationarity of the potential at the vacuum.  
The relevant equations are
\begin{align}
G_{\bar{T}}+G^{\bar{I}}\nabla_{\bar{T}}G_{\bar{I}}=3\delta \left( G_{\bar{T}}+\frac{2}{3}\right), \label{VTbarEq} \\
G_{\bar{S}}+G^{\bar{I}}\nabla_{\bar{S}}G_{\bar{I}}=3\delta \left(G_{\bar{S}}-\frac{2}{3}N_{\bar{S}} \right) \label{VSbarEq},
\end{align}
where $\delta=\frac{g^{2}}{6m_{3/2}^{2}}D^{A}D_{A}$ is the D-term SUSY breaking fraction.
More explicitly, eq.~\eqref{VTbarEq} is
\begin{align}
3\delta G_{\bar{T}} +2\delta =G_{\bar{T}}+G_{T}g^{\bar{I}T}\nabla_{\bar{T}}G_{\bar{I}}+G_{i}g^{\bar{J}i}\nabla_{\bar{T}}G_{\bar{J}}+G_{S}g^{\bar{I}S}\nabla_{\bar{T}}G_{\bar{I}}.
\end{align}
We concentrate on models that deviate only slightly from the simple model \eqref{simpleN}, \eqref{simpleF}, so we assume $|N_{S}|$ and $|F_{SS}|$ are at most of order $m_{3/2}/m_{\Phi}^{2}$.   We also use $S\simeq W/2$.
For example, 
\begin{align}
G_{Ti\bar{j}}=-\frac{3J_{i\bar{j}}}{(T+\bar{T}-N-J)^{2}}-\frac{6J_{i}J_{\bar{j}}}{(T+\bar{T}-N-J)^{3}}\simeq -\frac{1}{3}J_{i\bar{j}}.
\end{align}
Similarly, $\nabla_{T}G_{T}\simeq -2/3$, $\nabla_{T}G_{S}\simeq 2/W$, and $\nabla_{T}G_{i}\simeq -2G_{i}/3 $.
Equation~\eqref{VTbarEq} becomes
\begin{align}
0 \simeq (1-3\delta )G_{\bar{T}}+2\left(\frac{N^{\bar{S}}}{\bar{W}}-1\right)G_{T}-2+2G_{S}\frac{N^{\bar{S}S}}{\bar{W}}\simeq -2+2G_{S}\frac{N^{\bar{S}S}}{\bar{W}},
\end{align}
so $G_S$ is approximately given by
\begin{align}
G_{S}\simeq N_{S\bar{S}}\bar{W} \simeq \frac{12m_{3/2}}{m_{\Phi}^{2}}. 
\end{align}
This implies the tiny VEV of $T$:
\begin{align}
T\simeq \frac{3m_{3/2}^{2}}{m_{\Phi}^{2}}-\frac{F_{S}}{2}. \label{tshift}
\end{align}

In the same way, from eq.~\eqref{VSbarEq}, $\nabla_{S}G_{S}\simeq F_{SS}/W$ and $\nabla_{S}G_{i}\simeq -N_{S}G_{i}/3$, we obtain
\begin{align}
0 \simeq & \left(1-3\delta \right)G_{\bar{S}}+\left( \frac{2N^{S}+N^{\bar{S}S}\bar{F}_{\bar{S}\bar{S}}}{\bar{W}} \right)G_{S}+\left(\frac{6+N^{\bar{S}}\bar{F}_{\bar{S}\bar{S}}}{\bar{W}}\right)G_{T}+\frac{2J^{i}}{\bar{W}}G_{i}+2N_{\bar{S}}.
\end{align} 
To simplify the expression, let us assume $F_{SS}=\delta=0$, $N=-3+\frac{12}{m_{\Phi}^{2}}S\bar{S}$ with $S=W/2$ at the vacuum.
Then, $G_{T}$ becomes
\begin{align}
G_{T}\simeq -6\frac{m_{3/2}^{2}}{m_{\Phi}^{2}}-\frac{1}{3}J^{i}G_{i}.
\end{align}
If we further assume for the SUSY breaking sector that $J(z,\bar{z})=|z|^{2}-\frac{|z|^{4}}{\Lambda^{2}}$ and $P(z)=\mu^{2}z+W_{0}$,
$J^{i}$ is given by $J^{z}\simeq |z|\left( 1+ \frac{2}{\Lambda^{2}}|z|^{2}\right)\simeq |z| \simeq \sqrt{12}\left( \frac{m_{3/2}}{m_{z}} \right)^{2}$, where $m_{z}\simeq \sqrt{12}m_{3/2}/\Lambda$.
It is implied that
\begin{align}
S\simeq \frac{W}{2}\left(\frac{3}{A}+G_{T} \right)=\frac{W}{2}+\mc{O}\left( \frac{m_{3/2}^3}{m_{\Phi}^{3}} \, \text{or} \, \frac{m_{3/2}^{3}}{m_{\Phi}m_{z}^{2}}  \right). \label{sshift}
\end{align}
Equations~\eqref{tshift} and \eqref{sshift} can be used to obtain shifts of quantities {\it e.g.} $A\simeq 3+ 3m_{3/2}^{2}/m_{\Phi}^{2}$ induced by SUSY breaking.

The mixing matrix $\mc{A}$ 
 has two effects: canonicalization of kinetic terms and diagonalization of mass terms.
We assume that there is a single SUSY breaking field $\phi^{z}=z$, and its kinetic term and mass term are dominated by the diagonal part (proportional not $zz$ nor $\bar{z}\bar{z}$ but to $z\bar{z}$) for simplicity, and then the matrix element is simplified \cite{Endo:2012yg}
\begin{align}
(\mc{A}^{-1})^{z}{}_{\Phi_{\text{R}\pm}}=&\frac{g^{\bar{z}z}}{m_{\Phi}^{2}-m_{z}^{2}}  \left( \frac{\sqrt{3}}{2} \left( V_{T\bar{z}}+V_{\bar{T}\bar{z}}+J_{\bar{z}} \left(V_{T\bar{T}} +V_{\bar{T}\bar{T}}\right)\right)
 \pm \frac{m_{\Phi}}{4\sqrt{3}}    \left( V_{S\bar{z}}+V_{\bar{S}\bar{z}}+J_{\bar{z}} \left(V_{S\bar{T}} +V_{\bar{S}\bar{T}}\right)\right) \right) .
\end{align}
For the former part regarding $T$, only the $V_{T\bar{T}}\simeq 4N^{\bar{S}S}$ term remains.
If $m_{\Phi}^{2}\gg m_{z}^{2}$, this term cancels the term in $G_{T}$ proportional to $G_{z}$.
For the latter part regarding $S$, all the four terms are nonzero: 
\begin{align}
\tilde{V}_{TS}=&-8N^{\bar{S}S}N_{SS\bar{S}}|S|^{2}-4\bar{S}, \\
\tilde{V}_{T\bar{S}}=&-8N^{\bar{S}S}N_{S\bar{S}\bar{S}}|S|^{2}+2N^{\bar{S}S}\left(\bar{F}_{\bar{S}\bar{S}} +2N_{\bar{S}}+2\bar{S}N_{\bar{S}\bar{S}}\right)-8S, \\
\tilde{V}_{S\bar{z}}+\tilde{V}_{\bar{S}\bar{z}}=&\left(4 \left(S+\bar{S}\right)J^{\bar{k}}+2P_{l}J^{l\bar{k}}\right)\left( J_{\bar{k}\bar{z}}-J_{\bar{z}k\bar{k}}J^{k}\right) -4\bar{P}_{\bar{z}} +2\bar{P}_{\bar{k}\bar{z}}J^{\bar{k}} -2\bar{P}_{\bar{k}}J^{\bar{k}l}J_{l\bar{m}\bar{z}}J^{\bar{m}},
\end{align}
at the vacuum.
Among these, $-4\bar{P}_{\bar{z}}$ cancels the leading term in $G_{S}=12m_{3/2}/m_{\Phi} + \cdots$ under the same condition $m_{\Phi}^2 \gg m_{z}^{2}$.
Assuming $J(z,\bar{z})=|z|^{2}-\frac{|z|^{4}}{\Lambda^{2}}$ and $P(z)=\mu^{2}z+W_{0}$, subleading terms regarding this cancellation are still subdominant compared to terms in $G_{T}$.

In summary, the effective coupling is approximated as
\begin{align}
\left | \mathcal{G}^{\text{(eff)}}_{\Phi_{\text{R}\pm}} \right |^{2}\simeq & 2 \left| \frac{\sqrt{3}}{2} \left( -6\frac{m_{3/2}^{2}}{m_{\Phi}^{2}}+\frac{1}{3}J^{z}G_{z}\frac{m_{z}^{2}}{m_{\Phi}^{2}-m_{z}^{2}} \right) \pm\frac{m_{\Phi}}{4\sqrt{3}}\left( \frac{12\bar{W}}{m_{\Phi}^{2}}-\frac{4G^{z}G_{z}\bar{W}}{m_{\Phi}^{2}-m_{z}^{2}} \right) \right|^{2} \nonumber \\
\simeq & 6 \left| 3\frac{m_{3/2}^{2}}{m_{\Phi}^{2}}+\frac{m_{z}^{2}}{m_{z}^{2}-m_{\Phi}^{2}} \left( \frac{1}{6}J^{z}G_{z}\mp \frac{\bar{W}}{m_{\Phi}}  \right)  \right|^{2}.
\end{align}
Finally, the effective coupling is simplified when $m_{z}$ is in particular ranges:
\begin{align}
\left | \mathcal{G}^{\text{(eff)}}_{\Phi_{\text{R}\pm}} \right |^{2}\simeq \begin{cases}
96\left(\frac{m_{3/2}}{m_{\Phi}}\right)^{4} & \left( m_{z}^{2}\ll m_{\Phi}m_{3/2} \right) \\
6\left( \frac{m_{z}^{2}m_{3/2}}{m_{\Phi}^{3}}\right)^{2} & \left( 3m_{\Phi}m_{3/2}\ll m_{z}^{2} \ll m_{\Phi}^{2} \right) \\
6 \left( \frac{m_{3/2}}{m_{\Phi}}\right)^{2} & \left( m_{\Phi}^{2}\ll m_{z}^{2}\right)
\end{cases},
\end{align}
where we have assumed again $J(z,\bar{z})=|z|^{2}-\frac{|z|^{4}}{\Lambda^{2}}$ and $P(z)=\mu^{2}z+W_{0}$ to evaluate $J^{z}G_{z}$.
Therefore, the gravitino pair production rate is
\begin{align}
\Gamma(\Phi_{\text{R}\pm} \to \psi_{3/2}\psi_{3/2}) \simeq \frac{m_{\Phi}^{3}}{48\pi M_{\text{G}}^{2}} \times \begin{cases}
16 \left( \frac{m_{3/2}}{m_{\Phi}} \right)^{2} & \left( m_{z}^{2}\ll m_{\Phi}m_{3/2} \right) \\
\left( \frac{m_{z}}{m_{\Phi}}\right)^{4} & \left( 3m_{\Phi}m_{3/2}\ll m_{z}^{2} \ll m_{\Phi}^{2} \right) \\
1 & \left( m_{\Phi}^{2}\ll m_{z}^{2}\right)
\end{cases}.
\end{align}

\section{Constraints from gravitino abundance} \label{sec:cosmology}

We study gravitino abundance produced during and after reheating of the universe.
Gravitino is generated by various processes, (i) direct decay of the inflaton, (ii) scattering in the thermal bath created by the inflaton decay, (iii) decay of particles such as $\chi^{S}$ and $z$ produced by inflaton decay, and (iv) decay of coherent oscillation of SUSY breaking field $z$.
Similar analyses have been done in the literature, see Refs.~\cite{Nakayama:2012hy, Evans:2013nka, Nakayama:2014xca} and references therein.

As for direct decay of inflaton (i), we have derived various partial decay rates in the previous sections.
We assume no significant entropy dilution occurs after the reheating of the universe due to the inflaton decay.  Note that the SUSY breaking field $z$ decays dominantly into a pair of gravitinos, so that it does not produce entropy when it decays.
We parametrize the total decay rate of inflaton as
\begin{align}
\Gamma _{\text{tot}}=X\frac{m_{\Phi}^3}{M_{\text{G}}^2} ,\label{total_decay_rate}
\end{align}
where $X$ is defined by this equation.
Among various decay channels, there is a generic decay channel via the anomaly-induced process.
If we assume that this is the dominant mode, 
 then $X$ is expressed as $X=N_{\text{g}}\alpha^2 b_0^2 /768\pi^3$ where $b_0=3T_{G}-T_{R}$. 
The branching ratio of the gravitino pair production is 
\begin{align}
\text{Br}(\Phi_{\text{R}\pm}\rightarrow \psi_{3/2}\psi_{3/2})\simeq \frac{1}{48\pi X} \times \begin{cases}
16 \left( \frac{m_{3/2}}{m_{\Phi}} \right)^{2} & \left( m_{z}^{2}\ll m_{\Phi}m_{3/2} \right) \\
\left( \frac{m_{z}}{m_{\Phi}}\right)^{4} & \left( 3m_{\Phi}m_{3/2}\ll m_{z}^{2} \ll m_{\Phi}^{2} \right) \\
1 & \left( m_{\Phi}^{2}\ll m_{z}^{2}\right)
\end{cases}.
\end{align}

The gravitino yield $Y_{3/2}\equiv \frac{n_{3/2}}{s}$ where $n_{3/2}$ is gravitino number density and $s$ is entropy density, due to direct decay of inflaton is given by
\begin{align}
Y_{3/2}^{\text{(direct)}}=\frac{3T_{\text{R}}\text{Br}_{3/2}}{2m_{\Phi}},
\end{align}
where $\text{Br}_{3/2}$ is the branching ratio into a gravitino pair, and we define the reheating temperature $T_{\text{R}}$ as
\begin{align}
T_{\text{R}}=\left( \frac{90}{\pi^2 g_{*}(T_{\text{R}})} \right)^{\frac{1}{4}} \sqrt{M_{\text{G}}\Gamma_{\text{tot}}}.
\end{align}
The gravitino yield becomes
\begin{align}
Y_{3/2}^{\text{(direct)}}=\left( \frac{90}{\pi^2 g_{*}(T_{\text{R}})} \right)^{\frac{1}{4}} \frac{1}{32\pi} \sqrt{\frac{m_{\Phi}}{XM_{\text{G}}}} \times \begin{cases}
16 \left( \frac{m_{3/2}}{m_{\Phi}} \right)^{2} & \left( m_{z}^{2}\ll m_{\Phi}m_{3/2} \right) \\
\left( \frac{m_{z}}{m_{\Phi}}\right)^{4} & \left( 3m_{\Phi}m_{3/2}\ll m_{z}^{2} \ll m_{\Phi}^{2} \right) \\
1 & \left( m_{\Phi}^{2}\ll m_{z}^{2}\right)
\end{cases}.
\end{align}

The gravitino yield from thermal bath is known to be \cite{Bolz:2000fu, Pradler:2006qh, Pradler:2006hh, Rychkov:2007uq, Kohri:2005wn, Nakayama:2014xca}
\begin{align}
Y_{3/2}^{\text{(thermal)}}\simeq \begin{cases} \text{min}\left [  2 \times 10^{-12} \left( 1+\frac{m_{\tilde{\text{g}}^{2}}}{3m_{3/2}^{2}} \right) \left( \frac{T_{\text{R}}}{10^{10}\text{GeV}} \right) , \frac{0.42}{g_{*s(T_{3/2})}} \right ] &  (T_{\text{R}}\gtrsim m_{\text{SUSY}} ) \\ 0 &  (T_{\text{R}}\lesssim m_{\text{SUSY}} ) \end{cases},
\end{align}
where $m_{\tilde{\text{g}}}$ is the gaugino (gluino) mass at zero temperature, $m_{\text{SUSY}}$ is the typical soft SUSY breaking mass.  We take them as $m_{\text{SUSY}}=m_{3/2}$, and $m_{\tilde{\text{g}}}=2.8\times 10^{-2}m_{3/2}$ (for $m_{3/2}\geq 10^{4.5}$ GeV; anomaly mediation) or $m_{\tilde{\text{g}}}=m_{3/2}$ (for $m_{3/2} < 10^{4.5}$ GeV; gravity mediation).

The inflaton decays into matter particles, gravitino, and SUSY breaking field.
It also decays into other SUGRA sector particles ($T$, $S$, $\chi^T$, and $\chi^S$) if kinematically possible, but the rate should be highly suppressed by the phase space factor.  Even if the decay is possible, these particles  decay shortly after they are produced if there are Giudice-Masiero terms $|J_{ij}|\sim \mc{O}(1)$.  
Moreover, gravitino abundance from decay of these SUGRA sector particles $X$ will be multiply suppressed by tiny branching ratios of $\text{Br}(\Phi_{\text{R}\pm} \rightarrow X+\text{anything})$ and $\text{Br}(X \rightarrow \psi_{3/2}+\text{anything})$.
Therefore we neglect effects of these SUGRA sector particles, and consider only the SUSY breaking field $z$ for the process of the type (iii).

The SUSY breaking field $z$ is produced as particles by the decay of inflaton, and it decays dominantly into a pair of gravitinos when $m_{\Phi}>2 m_{z}\gg m_{3/2}$ because the partial decay rate into them is enhanced by a factor $(m_{z}/m_{3/2})^2$,~\cite{Endo:2006zj, Nakamura:2006uc}
\begin{align}
\Gamma ( z \rightarrow \psi_{3/2}\psi_{3/2} )= \frac{m_{z}^{5}}{96\pi m_{3/2}^{2} M_{\text{G}}^{2}}, 
\end{align}
 while partial decay rates of other channels are of order $\Gamma = \mc{O}(m_{z}^3/4\pi M_{\text{G}}^{2})$.

The gravitino yield as a decay product of particle
$z$, which in turn is created by decay of the inflaton,
leads to
\begin{align}
Y_{3/2}^{\text{(particle)}}=\frac{2n_{z}}{s}=\frac{3T_{\text{R}}}{m_{\Phi}}\text{Br}(\Phi_{\text{R}\pm} \rightarrow zz) = \frac{T_{\text{R}}m_{z}^{2}}{16\pi X m_{\Phi}^{3}}.
\end{align}

Finally we consider the process of the type (iv).
For matter fields, canonically normalized Hubble-induced mass is $\sqrt{2}H$.
This value is close to that for critical damping $3H/2$, so matter fields rapidly moves to the instantaneous minimum, which can be regarded as zero for our purpose.
The SUSY breaking field $z$ is also trapped near the origin until it decays at $H=H_{\text{D}}\simeq \Gamma ( z \rightarrow \psi_{3/2}\psi_{3/2} ) $ or until it starts coherent oscillation at $H=H_{\text{O}}\simeq m_{z}$.
Here and hereafter the subscripts R, D, and O refer to the time of reheating, decay of $z$, and beginning of coherent oscillation of $z$, respectively.
Assuming that the dominant channel is the model-independent anomaly-induced decay, $H_{\text{R}}\simeq 2.2$ GeV.
For definiteness, we assume eqs.~\eqref{ZKahler} and \eqref{Zsuper} for the SUSY breaking sector.

The VEV of $z$ is evaluated as $\vev{z}\simeq 2\sqrt{3}\left(\frac{m_{3/2}}{m_{z}}\right)^{2}$, and the energy density of coherent oscillation $z$ is
\begin{align}
\rho_{z,\text{field}}=m_{z}^{2}\vev{z}^2=\frac{12m_{3/2}^4}{m_{z}^2} \times \begin{cases} 1 & (H>H_{\text{O}}) \\  \left( \frac{a}{a_{\text{O}}} \right)^{-3} & (H_{\text{O}}>H)
\end{cases},
\end{align}
where $a$ is the cosmic scale factor.
The entropy density is
\begin{align}
s=\frac{4\rho}{3T}=\frac{4H^2}{T}.
\end{align}
The gravitino yield from coherent oscillation of $z$ is thus
\begin{align}
Y_{3/2}^{\text{(field)}}=\frac{2\rho_{z,\text{field}}}{m_{z}s} = \frac{ 6 m_{3/2}^4 T_{\text{R}} }{ m_{z}^5 } \hspace{20pt} (H_{\text{O}}>H_{\text{R}}>H_{\text{D}}) .
\end{align}
If the mass scale of $z$ is larger than the inflation scale, $2m_{z}>m_{\Phi}$, $z$ goes close to its VEV during inflation, and the above quantity $Y_{3/2}^{\text{(field)}}$ is further suppressed by a factor $(m_{\Phi}^2/2m_{z}^2)$.

So far, we have implicitly assumed the decay of $z$ occurs at last.  
If the decay of $z$ occurs between $H_{\text{O}}$ and $H_{\text{R}}$, the energy density of gravitinos generated by the decay of coherent oscillation of $z$ at the time of reheating is
\begin{align}
\rho_{3/2} =\frac{12m_{3/2}^4}{m_{z}^2}\left( \frac{a_{\text{D}}}{a_{\text{O}}} \right)^{-3} \left(  \frac{a_{\text{NR}}}{a_{\text{D}}} \right)^{-4}\left(  \frac{a_{\text{R}}}{a_{\text{NR}}} \right)^{-3},
\end{align}
where NR stands for the time when gravitino becomes non-relativistic, $H_{\text{NR}}=(m_{3/2}/m_{z})^{3/2}H_{D}$.  The gravitino yield is
\begin{align}
Y_{3/2}^{\text{(field)}}=\frac{6m_{3/2}^{5}T_{\text{R}}}{m_{z}^6}.
\end{align}
If $z$ decays even earlier than $H_{\text{O}}$, the gravitino energy density is given by
\begin{align}
\rho_{3/2}=\frac{12m_{3/2}^4}{m_{z}^2} \left(  \frac{a_{\text{NR}}}{a_{\text{D}}} \right)^{-4}\left(  \frac{a_{\text{R}}}{a_{\text{NR}}} \right)^{-3},
\end{align}
so the gravitino yield is
\begin{align}
Y_{3/2}^{\text{(field)}}=\frac{6m_{3/2}^5 T_{\text{R}}}{m_{z}^4 H_{\text{D}}^2}.
\end{align}

In summary, the gravitino yield from coherent $z$ field is given by
\begin{align}
Y_{3/2}^{\text{(field)}}=\begin{cases}
\frac{ 6 m_{3/2}^4 T_{\text{R}} }{ m_{z}^5 } & (H_{\text{O}}>H_{\text{R}}>H_{\text{D}}) \\
\frac{6m_{3/2}^{5}T_{\text{R}}}{m_{z}^6}  & (H_{\text{O}}>H_{\text{D}}>H_{\text{R}}) \\
\frac{6m_{3/2}^5 T_{\text{R}}}{m_{z}^4 H_{\text{D}}^2} & (H_{\text{D}}> H_{\text{O}}>H_{\text{R}}) 
\end{cases}.
\end{align}

Because we assume no entropy production after inflaton decay until gravitino decay, the denominators of every $Y_{3/2}$ are common, so the cosmologically relevant gravitino yield is the sum of all four terms, $Y_{3/2}^{\text{(total)}}= Y_{3/2}^{\text{(direct)}} +Y_{3/2}^{\text{(thermal)}} +Y_{3/2}^{\text{(particle)}} +Y_{3/2}^{\text{(field)}}$.

\begin{figure}[htbp]
  \begin{center}
    \includegraphics[clip,width=7cm]{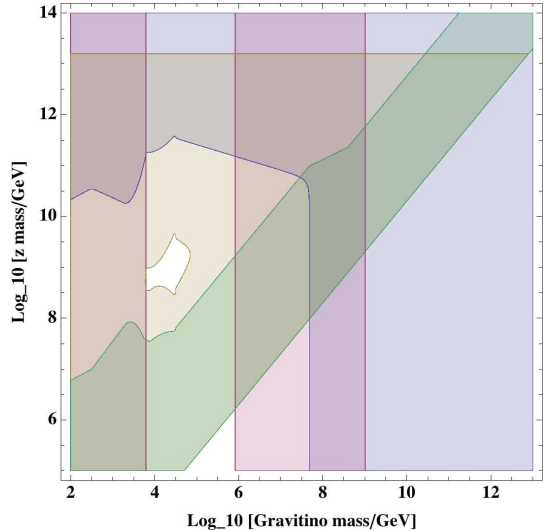}
    \caption{Constraint on masses of gravitino and SUSY breaking field from LSP overabundance from gravitino decay.  Blue, red, yellow, and green shaded regions, corresponding to direct production, thermal production, $z$ particle decay, and $z$ coherent oscillation decay, are excluded.}
    \label{fig:constraints}
  \end{center}
\end{figure}

Now that we have derived generic expressions for gravitino abundance, let us discuss its cosmological consequences for a minimal setup.
Gravitinos heavier than about $30$ TeV decay before big bang nucleosynthesis (BBN), but lightest supersymmetric particles (LSPs) produced by the gravitino decay chain may exceed the observed dark matter abundance.  Such a constraint is shown in Fig.~\ref{fig:constraints} assuming wino LSP of anomaly mediation for $m_{3/2} \geq 10^{4.5} \text{GeV}$.
For smaller gravitino mass, $m_{3/2} < 10^{4.5}\text{GeV}$, gravitino decay affects light element abundance.  We assume gravity mediation for this mass region and impose the standard BBN constraints~\cite{Kohri:2005wn} on the parameter space in Fig.~\ref{fig:constraints}. 
In this figure, the dominant decay mode of the inflaton is assumed to be a model-independent one, namely the anomaly-induced decay into gauge bosons and gauginos as discussed in subsection~\ref{sec:anomaly}.  The inflaton mass is taken as $m_{\Phi}=3.2 \times 10^{13}$ GeV, and the reheating temperature after inflaton decay is $T_{\text{R}}\simeq1.0\times 10^9$ GeV.
Instantaneous reheating occurs in spite of the Planck-suppressed interaction~\cite{Harigaya:2013vwa}.
 As can be seen from the Figure, most of the parameter space are excluded.
The lower unshaded region is also excluded by the standard constraint of the cosmological moduli problem~\cite{Coughlan:1983ci, Ref:Moduli} unless baryon asymmetry is regenerated \textit{e.g.}~by the Affleck-Dine mechanism~\cite{Affleck:1984fy}.  (In this case the modulus (Polonyi) field is the SUSY breaking field $z$.) 
Note that the range of gravitino mass $10^{6}\text{GeV} \lesssim m_{3/2} \lesssim 3\times10^{11}\text{GeV}$ (corresponding to $3 \text{TeV} \lesssim m_{\text{wino}}  \lesssim T_{\text{R}}$; not shown in the Figure) is excluded by thermally produced wino abundance~\cite{Hisano:2006nn} even without considering the wino LSP from gravitino decay.  
 See also Ref.~\cite{Moroi:2013sla} for non-thermal production of wino dark matter via the decay of long-lived particles.
 As usual, this problem is ameliorated or solved by assuming $R$-parity breaking so that LSP decays or thermal inflation~\cite{Lyth:1995ka} so that it is diluted.

\section{Summary and Discussion} \label{sec:conclusion}
In this paper, we studied coupling of the SUSY Starobinsky model to matter sector in the old-minimal supergravity, inflaton decay and its cosmological consequences.
To this end, we first transformed the supergravity theory of supercurvature $\mc{R}$ minimally coupled to matter to an equivalent one in the form of the standard no-scale type supergravity of inflaton $T$ plus another matter superfield $S$.
The notable feature there is that the interactions of the inflaton $T$ to other superfields in the theory are completely determined by the fact that the origin of $T$ is a Lagrange multiplier.
In particular, the inflaton $T$ does not enter in the gauge kinetic function.
These are characteristic features of the SUSY Starobinsky model, unlike some other SUGRA models having Starobinsky-like scalar potentials.

On the other hand, interactions of $S$ have more freedom.  In this paper, we assumed minimal coupling between SUGRA sector and matter sector in the first place, but it is not protected by any symmetries so more general coupling between $S$ and matter are possible.  It may enhance decay rates of inflaton into matter through mixing between $T$ and $S$, which results in a suppressed branching ratio into gravitino.
 
 We focused on model-independent decay channel of inflaton into gauge sector via the anomaly-induced decay in section~\ref{sec:cosmology}, but presence of heavy matter, like right-handed (s)neutrinos, and large quadratic holomorphic term $J_{ij}$ in K\"{a}hler potential, which is used for the Giudice-Masiero mechanism~\cite{Giudice:1988yz}, are helpful to reheat the universe efficiently.  These are simply because there are decay modes whose rates are proportional to matter mass or $J_{ij}$.
 
 Taking anomaly-induced decay into the gauge sector as the dominant decay channel, the lower limit of the reheating temperature is a similar value, $T_{\text{R}}\simeq 1.0\times 10^{9}$ GeV, to that of the non-SUSY original Starobinsky model, and it is consistent with thermal leptogenesis~\cite{Fukugita:1986hr}.
 The most striking difference to the non-SUSY case is presence of the built-in long lived particle in the theory, gravitino.
We assumed gravity/anomaly mediation of SUSY breaking, and estimated the amount of LSPs produced from decay of gravitino, which is produced either by direct decay of inflaton, thermal scattering, decay of SUSY breaking particle or field $z$.
The result is that most of the parameter space $(m_{3/2}, m_{z})$ is excluded unless $R$-parity is broken or thermal inflation occurs.
Thus, our prediction of the mass of gravitino is $10^{4} \text{GeV} \lesssim m_{3/2} \lesssim 10^{5} \text{GeV}$.
A way around this is considering more general coupling between SUGRA sector and matter sector in the original higher supercurvature SUGRA theory.

\section*{Acknowledgements}

We would like to thank Iannis Dalianis for clarifying the model in Ref.~\cite{Dalianis:2014aya}.
TT thanks Motoi Endo, Kazunori Nakayama, Fuminobu Takahashi, and Masahiro Takimoto for valuable discussion.
TT was supported partly by a grant of Advanced Leading Graduate Course for Photon Science in the University of Tokyo, and partly by a Grant-in-Aid for JSPS Fellows, and a Grant-in-Aid of the JSPS under No.~26$\cdot$10619.
YW acknowledges supports from the JSPS Research Fellowship for Young Scientists No.~269337 and the Munich Institute for Astro- and Particle Physics (MIAPP) of the DFG cluster of excellence ``Origin and Structure of the Universe."
YY acknowledges support from the JSPS Research Fellowship for Young Scientists No.~264236.
JY acknowledges support form the JSPS Grant-in-Aid for Scientific Research (B) No.~23340058.

\appendix
\section{Duality transformation of higher derivative SUGRA models }\label{app}
In this Appendix, we briefly review the duality transformation between the higher derivative SUGRA system and the standard one described in Sec.~{\ref{sec:setup}.  We explicitly show that the superpotential and the gauge kinetic function in the standard SUGRA are linear in and independent of $T$, respectively. 
We also discuss some generalizations.

The action including general couplings between ${\cal R}$ and matters~\cite{Cecotti:1987sa} is given by 
\begin{align}
S=&\int d^4xd^4\theta EN\left({\cal R},\bar{\cal R},\phi,\bar{\phi}e^{gV}\right)\nonumber\\
&+\left[\int d^4x d^2\Theta 2\ms{E}\left(F\left({\cal R},\phi \right)+\frac{1}{4} h_{AB}\left({\cal R},\phi \right)W^AW^B\right)+{\rm H.c.}\right ],\label{genR2}
\end{align}
where ${\cal R}$, $\phi$ and $W^A$ are the same as in Sec.~\ref{sec:setup}, $N$ is a real function of ${\cal R}$, $\phi$ and their conjugates, and $F$ and $h_{AB}$ are holomorphic functions of ${\cal R}$ and $\phi$. By introducing the Lagrange multiplier chiral multiplet $T$ and a chiral multiplet $S$, the action~(\ref{genR2}) becomes the following form,
\begin{align}
S=&\int d^4xd^4\theta EN\left(S,\bar{S},\phi,\bar{\phi}e^{gV}\right)\nonumber\\
&+\left[\int d^4x d^2\Theta 2\ms{E}\left(2T(S-{\cal R})+F(S,\phi )+\frac{1}{4}h_{AB}(S,\phi)W^AW^B \right)+{\rm H.c.}\right ].\label{multiplier}
\end{align}
Varying it with respect to $T$ yields the equation ${\cal R}=S$, and we obtain the original action~(\ref{genR2}). We can also rewrite the action~(\ref{multiplier}) into the standard SUGRA form not containing higher curvature terms as 
\begin{align}
S=&\int d^4xd^4\theta E\left[ N\left(S,\bar{S},\phi,\bar{\phi}e^{gV} \right)-\left(T+\bar{T}\right)\right] \nonumber\\
&+\left[\int d^4x d^2\Theta 2\ms{E}\left( 2TS+ F(S,\phi )+\frac{1}{4}h_{AB}(S,\phi)W^AW^B\right)+{\rm H.c.}\right ]\nonumber\\
=&\int d^4xd^2\Theta 2\ms{E}\left[\frac{3}{8}\left(\bar{\ms{D}}\bar{\ms{D}}-8{\cal R}\right)e^{-K/3}+W+\frac{1}{4}h_{AB}(S,\phi)W^AW^B\right]+{\rm H.c.},\label{dual}
\end{align}
where
\begin{align}
K&=-3\ln\left(\frac{T+\bar{T}-N\left(S,\bar{S},\phi,\bar{\phi}e^{gV}\right)}{3}\right),\\
W&=2TS+F(S,\phi).
\end{align}
Notice that in the dual action~(\ref{dual}), $T$ does not appear in the gauge kinetic function $h_{AB}$ even if there are couplings between ${\cal R}$ and $W^A_{\alpha}$ in the original action~(\ref{genR2}). The absence of $T$ in the gauge kinetic function $h_{AB}$ is a remarkable feature of the Starobinsky inflation in old-minimal SUGRA models, which restricts the main reheating processes to the anomaly induced decays into the gauge sector as discussed in Sec.~\ref{sec:decay}.

We have shown that the naive generalization of the action~(\ref{minimal}) does not contain $T$-dependence in gauge kinetic function $h_{AB}$. 
One may wonder what happens if we introduce dependence of the gauge kinetic function on $T$ in eq.~\eqref{dual} and transform it back to a higher derivative SUGRA.
As a minimal extension of eq.~\eqref{dual}, let us consider the following action in which the gauge kinetic function linearly depends on $T$,
\begin{align}
S=&\int d^4xd^4\theta E\left[ N\left(S,\bar{S},\phi,\bar{\phi}e^{gV}\right)-\left(T+\bar{T}\right)\right]\nonumber\\
&+\left[\int d^4x d^2\Theta 2\ms{E}\left( 2TS+F(S,\phi )+\left(\frac{1}{4}h_{AB}(S,\phi)-2H_{AB}T\right)W^AW^B\right)+{\rm H.c.}\right ],\label{ext}
\end{align}
where $H_{AB}$ is a constant. Here, to obtain the dual action of (\ref{ext}), we follow the way discussed in Ref.~\cite{Cecotti:2014ipa}. We can recast the action~(\ref{ext}) into the dual form as
\begin{align}
S=&\int d^4xd^4\theta E N\left(S,\bar{S},\phi,\bar{\phi}e^{gV}\right)\nonumber\\
&+\left[\int d^4x d^2\Theta 2\ms{E}\left(2T \left(S-{\cal R}-H_{AB}W^AW^B \right)+F(S,\phi )+\frac{1}{4}h_{AB}(S,\phi)W^AW^B\right)+{\rm H.c.}\right ].
\end{align}
Varying the above action with respect to $T$ yields $S={\cal R}+H_{AB}W^AW^B$. Substituting it into the action gives 
\begin{align}
S=&\int d^4xd^4\theta E N\left({\cal R}+H_{AB}W^A W^B,\bar{\cal R}+\bar{H}_{AB}\bar{W}^A \bar{W}^B,\phi,\bar{\phi}e^{gV}\right) \nonumber\\
&+\left[\int d^4x d^2\Theta 2\ms{E}\left(F\left({\cal R}+H_{AB}W^AW^B,\phi \right)+\frac{1}{4}h_{AB}\left({\cal R}+H_{CD}W^CW^D,\phi \right) W^AW^B \right)+{\rm H.c.}\right ].\label{exdual}
\end{align}
Notice that the dual action~(\ref{exdual}) contains higher dimensional operators involving $H_{AB}W^AW^B$. It means that the theory contains the higher derivative terms of the gauge multiplets $V^A$. 
In such a case, the inflaton $T$ can decay into gauge bosons and gauginos through tree level couplings in the gauge kinetic function (or in the K\"{a}hler potential depending on the shift of $S$). Then, reheating processes can be different from the ones we discussed in Sec.~\ref{sec:decay}.

For completeness, we finally discuss a possibility that the superpotential and the gauge kinetic function are non-linear in $T$.
We generalize the $F$-term action as 
\begin{align}
S=&\int d^4xd^4\theta E\left[ N\left(S,\bar{S},\phi , \bar{\phi}e^{gV}\right)-\left(T+\bar{T}\right)\right]\nonumber\\
&+\left[\int d^4x d^2\Theta 2\ms{E}\left(F(T, S, \phi )+\frac{1}{4}h_{AB}(T, S, \phi) W^{A} W^{B}\right)+{\rm H.c.}\right ],\label{nl}
\end{align}
where $F(T, S, \phi )$ and $h_{AB}(T, S, \phi )$  are holomorphic functions of $T$, $S$, and $\phi^i$. We can rewrite the action~(\ref{nl}) as 
\begin{align}
S=&\int d^4xd^4\theta EN\left(S,\bar{S},\phi , \bar{\phi}e^{gV}\right)\nonumber\\
&+\left[\int d^4x d^2\Theta 2\ms{E}\left(-2\mc{R}+F(T, S, \phi )+\frac{1}{4}h_{AB}(T, S, \phi) W^{A} W^{B}\right)+{\rm H.c.}\right ]. \label{nldual}
\end{align}
Varying the above action with respect to $T$ yields
\begin{align}
F_{T} (T, S, \phi )+ \frac{1}{4}h_{AB, T} (T, S, \phi ) W^{A}W^{B} - 2\mc{R}=0,
\end{align}
and it can be implicitly solved as $S=S(\mc{R}, T, \phi )$.
Substituting it to eq.~\eqref{nl} leads to a higher derivative SUGRA depending on $\mc{R}$, $\phi^i$, and an additional matter $T$. 
Notice that dependence on the additional matter $T$ vanishes  if and only if $F(T, S, \phi )$ and $h_{AB}(T,S, \phi )$ are linear functions of $T$. Therefore, the non-linear dependence of $T$ in the superpotential or the gauge kinetic function requires a new chiral multiplet $T$ in the dual higher derivative SUGRA theory.

In this paper, we focus on the case that $T$ appears as the degree of freedom purely originated from the higher derivative SUGRA terms.
In this case, the corresponding action of the standard SUGRA is given by eq.~\eqref{TransformedTheory} [or more generally by eq.~\eqref{dual}].}

\end{document}